\documentstyle[12pt]{article}
\newcommand{\be}{\begin{equation}}
\newcommand{\ee}{\end{equation}}

\begin{document}

\begin{titlepage}
\centerline{\large\bf INSTITUTE OF THEORETICAL AND EXPERIMENTAL PHYSICS}
\vspace{7cm}
\centerline{\large\bf B.Kerbikov}
 \vspace{2cm}
  \centerline{\large\bf
 THE INTERPLAY OF THE  $K^+K^-$ ATOM}
 \vspace{5mm}
\centerline{\large\bf AND THE $f_0(975)$ RESONANCE}
\vspace{5cm}

\centerline{\large\bf Moscow - 1995}
\vspace{1cm}
$~~$

\end{titlepage}

\newpage
$~~$
\vspace{60mm}

\centerline{\large\bf A b s t r a c t}
\vspace{5mm}
\large

We predict that production of the
$K^+K^-$ atom  in $pd\to ~^3He~X$ and  similar reactions exhibits a drastic
missing mass spectrum due to the interplay with $f_0(975)$ resonance. We
point out that high precision studies of the $K^+K^-$ atom may shed a new
light on the nature of $f_0(975)$.

\newpage
\vspace{1cm}

Experiments aimed at the detection of the $K^+K^-$ atom (kaonium)
are now being contemplated (see [1] and references
therein). In the present paper we wish to provide a sound motivation for
kaonium searches. The underlying physics of our approach is quite different
from that of the pioneering paper [1] on two--meson atoms. Our main emphasis
will be on the interplay of the kaonium and the $f_0(975)$ meson which lies
in the intermediate vicinity of the $K^+K^-$ threshold. The nature of $f_0$
is a long standing problem (see, e.g.[2-5]). It will be shown that kaonium may
be an additional  eye-witness of the $f_0$ nature.

To be specific we shall consider the production of kaonium in the reaction
\be
pd\to~^3He~X\to~^3He~~\pi\pi
\ee
with the invariant mass of $X$ being in the
vicinity of the $K^+K^-$ threshold although the same formalism can be
applied also to other kaonium production reactions.
Kaonium  has the Bohr radius of
$a_B=1/\alpha\mu = 109.6 fm$, the ground state binding energy is
$E_0=\alpha^2\mu/2=6.57 keV$, where $\mu=m_K/2$. The expected decay width is
$\Gamma= 4\alpha^3\mu^2{\rm Im}~ A\simeq 0.5 keV$ [1]. This is a factor
$10^{-5}$ smaller than the width of the $f_0(975)$ meson. Therefore
following the arguments of doorway states theory in nuclear physics [6], or
the theory of ``zombie-levels''in particle physics [6-10], we shall assume
that kaonium decay proceeds only via its mixing with $f_0$.  Mixing of
kaonium with $f_0$ (and $a_0(980)$ -- see below) is what makes kaonium
production and decay different from other hadronic atoms,  including not
only pionium $\pi^+\pi^-$ [1,11] but also kaonic hydrogen $K^-p$ for which a
resonance $\Lambda(1405)$ plays  a certain role [12]. It is clear that the
mixing parameter of  $f_0$ and kaonium is  proportional to the atomic wave
function at the origin. On top of this trivial geometrical dependence the
mixing parameter is, as we shall see, sensetive to the nature of $f_0$.

There are several ways to write a unitary amplitude describing the mixture
of a broad and a narrow resonances under the assumption that the width of
the narrow one is induced by its mixing with the broad one. In this short
note we shall make use  of the mass-matrix formalism [7]. The mass-matrix
of our system reads
 \begin{equation}
  H-E= \left( \begin{array}{cc}
 E_r-i\displaystyle{\frac{\Gamma_r}{2}}-E& \gamma\\
 \gamma&E_a-E
 \end{array} \right ).
  \end{equation}

The first level corresponds to $f_0$ so that $E_r=975MeV,~~\Gamma_r=50
MeV$, the second level is kaonium with $E_a=2m_K-6.57 keV,~~\gamma$ is the
mixing parameter which can be taken real [7] and which will be estimated
below.
 We note in passing that the basis behing Eq. (2) is not orthogonal since
 the matrix $H$ is not Hermitian. Next we make the assumption that  reaction
(1) close to
to $K^+K^-$ threshold is dominated by $f_0(975)$ production, i.e.  there is
no direct production of the $K^+K^-$ atom. Let $M_r$ be the amplitude of
$f_0$ production in reaction (1), $\Gamma_{\pi}$ be the decay width
$f_0\to \pi\pi$.
We remind that the state $X$ in (1) is a mixture of $f_0$ and kaonium
corresponding to the mass-matrix (2). Then the amplitude for reaction (1) is
given by [7]
\be
T=\sqrt{\Gamma_{\pi}}\left(\frac{1}{H-E}\right)_{rr}M_r=
\frac{\sqrt{\Gamma_{\pi}}(E-E_a)M_r}{(E_a-E)(E_r-i\displaystyle
{\frac{\Gamma_r}{2}}-E)-\gamma^2}.
\ee
The stricking feature of this amplitude is that is has a zero at $E=E_a$,
i.e. at the kaonium position (compare to the well known ``dipole'' phenomena
considered in chapter 8 of [13]). Eq. (3) can be manipulated into a sum of
two Breit-Wigner amplitudes. In the approximation $\gamma \ll
|E_r-E_a-i\displaystyle{\frac{\Gamma_r}{2}}| $ we get
\be
T\simeq\left[
\frac{\sqrt{\Gamma_{\pi}}}
{E-E_r+i\displaystyle{\frac{\Gamma_r}{2}}}
+\frac{1}{(E-E'_{a}+
i\displaystyle{\frac{\Gamma_a}{2})}}
\frac{\gamma^2\sqrt{\Gamma_{\pi}}}{(E_a-E_r+i
\displaystyle{\frac{\Gamma_r}{2}})^2}\right]M_r,
\ee
where
\be
\Gamma_a=\frac{\gamma^2\Gamma_r}{(E_r-E_a)^2+
\displaystyle{\frac{\Gamma_r^2}{4}}},~~
E_{a}'=E_a+\gamma^2\frac{E_a-E_r}{(E_r-E_a)^2+
\displaystyle{\frac{\Gamma_r^2}{4}}}.
\ee
Eq. (4) reflects the fact that the mass matrix (2) has two eigenvalues, one
of which is close to the ``bare'' $f_0$ eigenvalue, while the other
corresponds to the atomic state which has acquired some shift and width due
to the mixing with $f_0$. Also Eq. (4) clearly exhibits the zero--peak
structure of the amplitude $T$ in the vicinity of $E_a$.

Using the standard representation of the 3-body phase--space [14] it is easy
to show  that the missing mass distribution in reaction (1) is given by
\be
\frac{d\sigma}{dm^2_X}=C|T(E)|^2,
\ee
where $C$ (as well as $M_r$) is practically energy independent within the
$m_X$ interval of the order of $\Gamma_r$ around the
$f_0$ position.

Now the point at  issue is to estimate the mixing parameter $\gamma$
which governs the scale of the zero-peak structure of $ d\sigma/dm^2_X$. This
is where the nature of $f_0$ comes into play. Consider first the
interpretation of $f_0$ as $K\bar{K}$ molecule, i.e. a deuteron-like state.
Then introducting effective transition operator $V$, we get the following
estimate
$$
\gamma =<\psi_a|V|\psi_r>=\int
d{\bf {p}}<\psi_a|{\bf {p}}><{\bf{p}}|V|\psi_r>\simeq
$$
$$
\simeq (2\pi)^{3/2}<{\bf{p}}=0|V|\psi_r>\psi_a({\bf{r}}=0)
\simeq - (2\pi)^{3/2}\psi_a(0)\frac{(m_K\varepsilon)^{1/4}}{\pi m_K},
$$
where $\varepsilon \simeq 12 MeV$ is the binding energy of the
$K\bar{K}$ molecule.
Thus
\be
\gamma^2\simeq
\frac{8\pi}{m_K}(\psi(0))^2(\frac{\varepsilon}{m_K})^{1/2}
=\alpha^3m_K(m_K\varepsilon)^{1/2} \simeq 2\cdot 10^{-2}MeV^2.
\ee

For the alternative compositions of  $f_0$ ( e.g. $q\bar{q}$,
multiquark, glueball, etc.), the natural way to parametrize $\gamma^2$ is
through the Jaffe--Low $P$ --matrix [15].
In $P$ -- matrix terminology  $f_0$ is the $P$--matrix ``primitive''  with
eigenvalue $E_n$, radius $b$, and the coupling to hadronic channels given by
a  residue $\lambda_n$. In terms of these  quantities the $P$ -- matrix
reads:
 $$P=k~cot(kb+\delta)=P_0+\frac{\lambda_n}{E-E_n}.$$

   Straightforward  calculations [10,16] lead to the
  estimate \be \gamma^2= \frac{8\pi}{m_K}\lambda_n b^2(\psi(0))^2
=\alpha^3m_K^2\lambda_nb^2.
\ee
 The value of the residue $\lambda_n$ of $f_0$
into $K^+K^-$ channel is subject to large uncertainties. As an educated
guess we can take the value of $\lambda_n$ for $q^2\bar{q}^2$ states in 1
Gev mass region from [17]. This yields
$\gamma^2\sim 10^{-2} MeV^2$ for $b=0.8 fm$
and to $\gamma^2\sim5\cdot10^{-4} ~MeV^2$ for
$b=0.2 fm$ as suggested in [5].

Another argument that the estimate $\gamma^2\sim 10^{-2} MeV$  is reliable
follows directly from our Eq.(5) for the induced width of kaonium. This
equation shows that the expected [1] value of kaonium width  $\Gamma_a\sim
0.5 keV$ corresponds to  $\gamma^2\sim 10^{-2} MeV^2 $. It is a sound
confirmation that our general scheme given by Eqs.(2-5) is self-consistent.
For $\gamma^2= 10^{-2} MeV^2 $  the separation in energy between the zero
and the peak in $d\sigma/dm^2_X$ distribution is $2 keV$.
This is still much less than the distance to the first excited state in
 kaonium.
 In Fig. 1 we plot the $d\sigma/dm^2_X$ distribution in arbitrary units
normalized to 1 at $E=E_r$.

 In conclusion, we would like to add some remarks on the details  and
complications glossed over in the preceding discussion, and on the
comparison of the above results with that of Ref. [1]. The first point
to improve  on is the inclusion of the background.
In a paper to follow the background is taken into account using the
$T$-- matrix formalism  in line with Stodolsky's approach to resonance
mixtures [8]. The main source of the background is the  direct (not via
$f_0$) production of the $\pi\pi$ system. To our knowledge the excitation
function of the $f_0$ meson in $pd\to~^3HeX$ reaction is at present not
available [18]. Therefore the accurate estimate of the background is
impossible.  Unless background becomes dominant  the pattern of the cross
section outlined
above is still present. However instead of  vanishing at the position of the
atomic level the cross section undergoes a deep  minimum shifted with
respect to the atomic eigenvalue. The $ a_0$ meson also remained out of
the scope of the present paper. The decay of the $a_0$ meson into
$\pi\pi$ is forbidden, but kaonium is mixed with both $f_0$ and $a_0$
mesons. The physics of  the ($f_0- a_0$--kaonium) system deserves
special discussion.
Here we only mention that the distortion  of  the above picture due to
$a_0$ depends on the relative phase of $f_0$ and $a_0$ production in
$pd$ collisions. The last subtlety of our treatment is the constant
width Breit--Wigner approximation for the $f_0$. The whole  phenomenon
under consideration is taking place in a very narrow energy interval of
the order of 10 keV around  $K^+K^-$ threshold. Therefore the dominant
partial width $\Gamma_{\pi}$ of $f_0$ into $\pi\pi$ channel is really
constant. The same   is true for the $f_0$ production amplitude $M_r$
(see Eq. (3)). Looking back at Eq. (3)  we see that constant
$\Gamma_{\pi}$ and $M_r$ are at the core  of the observated structure.
Concerning the $f_0\to K\bar{K}$ partial width the situation is the
following.  The threshold  of the $K^0\bar{K}^0$ channel is 7 MeV above
the atomic energy and therefore the corresponding partial width also can
be considered constant in a narrow energy  interval under consideration. The
$f_0\to K^+K^-$ partial width displays the structure typical for the
opening of the oppositely charged particles  channel.
However this structure is not relevant for the interplay of the $f_0$ meson
and kaonium ground state. It is important for the cusp at the $K^+K^-$
threshold and /or for the formation of the kaonium excited states.

Some remarks are due on the comparison with Ref. [1].
Our main assumption is the dominance of $f_0$ meson in kaonium production.
In Ref. [1] such  a  situation is only mentioned in passing while the
position and the width of $f_0$ enter in no equation whatsoever. Direct
kaonium production (not via $f_0$)  considered in [1] plays the role of a
background in our interpretation. Zero--peak structure due to the interplay
of $f_0$ and kaonium is intrinsic to our mechanism. In [1] there is also a
contribution which exhibits similar behavior (so--called Fano resonance
[19]). The ``Fano zero'' is due to the interference in the elastic channel
of a smooth background and a Breit--Winger amplitude [20]. For reaction (1)
Fano structure appears only due to the process
$
pd\to~^3He~\pi\pi \to ~^3He(K^+K^-)_{Atom}\to~^3He~\pi\pi.
$

In our treatment this reaction  is just a part of a
background. To conclude the comparison we may say that the present approach
is complementary to that of Ref. [1]. To answer the question which of the
two mechanisms is dominant one needs an accurate experimental study of the
$K^+K^-$ threshold region in reaction (1).
We can hope that our paper succeeds in motivating experimental searches for
the kaonium. We  also hope that it can provide the  experimentalist with
useful advice on how to find this elusive yet fascinating object.

It is a pleasure to thank  S. Bashinsky, M.Voloshin, K.Boreskov and
N.Demchuk for useful discussions. I am indebted to A.Green for providing me
with their paper [1] prior to publication and to L.Nemenov for focusing my
attention on the problem.
  \newpage
   \begin{center}

{\large Figure capture}\\
\end{center}

Fig.1  Missing mass distribution $d\sigma/dm^2_X$ in reaction (1) in
arbitrary units.
\newpage

\begin{center}

{\large REFERENCES}\\

\end{center}
\begin{enumerate}

\item S.Wycech, A.M.Green, Nucl. Phys. {\bf A562}, 446 (1993).
\item J.Weinstein and N.Isgur, Phys. Rev. {\bf D41}, 2236 (1990).
\item D.Morgan and M.R.Pennington, Phys. Rev. {\bf D48}, 1185 (1993).
\item B.S.Zou and D.V.Bugg, Phys. Rev. {\bf D48}, R3948(1993).
\item .E.Close et al. Phys. Lett. {\bf B319}, 291 (1993).
\item B.Block, H.Feshbach, Ann. Phys.{\bf 23}, 47 (1963).
\item I.Kobzarev, N.Nikolaev, L.Okun, Yad. Fiz. {\bf 10}, 864 (1969).
\item L.Stodolsky, Phys.Rev. {\bf D1}, 2683 (1970).
\item A.Kudryavtsev, Yad. Fiz. {\bf 10}, 309 (1969).
\item B.Kerbikov, Theor. and Math. Phys. {\bf 65}, 1225 (1985).
\item L.Nemenov, Sov. J. Nucl. Phys. {\bf 41}, 629 (1985).
\item K.Tanaka, A.Suzuki, Phys. Rev. {\bf C45}, 2068 (1992).
\item M.L.Goldberger and K.M.Watson, Collision Theory, (John Wiley, N.Y. 1964).
\item E.Byckling and K.Kajantie, Particle Kinematics, (John Wiley, N.Y. 1973).
\item R.L.Jaffe and F.E.Low, Phys. Rev. {\bf D19}, 2105 (1979).
\item B.L.G.Bakker et al. Sov. J. Nucl. {\bf 43}, 982 (1986).
\item R.P.Bickerstaff, Phil. Trans. R. Soc. Lond. {\bf A309}, 611 (1983).
\item A.Codino and F.Plouin, Production of light mesons and multipion
systems in light nuclei interactions. Preprint LNS /Ph/
94-06.
\item U.Fano, Phys. Rev. {\bf 124}, 1866 (1961).
\item J.M.Blatt and V.F.Weisskopf, Theoretical Nuclear
Physics, (John Wiley, New York, 1952), Ch. VII.
\end{enumerate}
\end{document}